\documentclass[useAMS,usenatbib,usegraphicx]{mn2e}

\title[Satellite lines and the He-like K$\alpha$ complex]{On the importance of satellite lines to the He-like K$\balpha$ complex and the {\boldmath $G$} ratio for calcium, iron, and nickel.}
\author[J. Oelgoetz et al.]
  {J.~Oelgoetz$^1$\thanks{Email: oelgoetzj@apsu.edu}\thanks{Present address: Department of Physics and Astronomy, Austin Peay State University, P.O. Box 4608, Clarksville, TN 37044, USA}
  C.~J.~Fontes$^1$, H.~L.~Zhang$^1$, S.~N.~Nahar$^2$, and A.~K.~Pradhan$^2$ \\
  $^1$Applied Physics Division, Los Alamos National Laboratory, Box 1663, MS F663, Los Alamos, NM 87545, USA \\
  $^2$Department of Astronomy, The Ohio State University, 140 W. 18th Avenue, Columbus, OH 43210, USA} 
\date{Released 2008 Xxxxx XX}

\pagerange{\pageref{firstpage}--\pageref{lastpage}} \pubyear{2008}
\usepackage[dvips,dvipsnames]{color}
\usepackage{amssymb}
\newcommand{\apj}{ApJ}

\newcommand{\aap}{A\&A}
\newcommand{\aaps}{A\&A Sup. Ser.}
\newcommand{\mnras}{MNRAS}
\newcommand{\pra}{PRA}

\newcommand{\apjs}{ApJ Sup. Ser.}

\newcommand{\jpbo}{J. {P}hys. {B}, {A}t., {M}ol. {P}hys.}

\newcommand{\nat}{Nat.}
\newcommand{\jqsrt}{J. Quant. Spectrosc. Radiat. Transf.}

\newcommand{\prsa}{Proc. Roy. Soc. Lon. A}
\begin{document}
\newlength{\figspace}
\setlength{\figspace}{0.15in}
\newlength{\figheight}
\setlength{\figheight}{0.78\textheight}
\newlength{\sfigheight}
\setlength{\sfigheight}{0.90\textheight}
\setcounter{page}{0}
\thispagestyle{empty}
\begin{center}
\resizebox{\textwidth}{!}{\includegraphics*{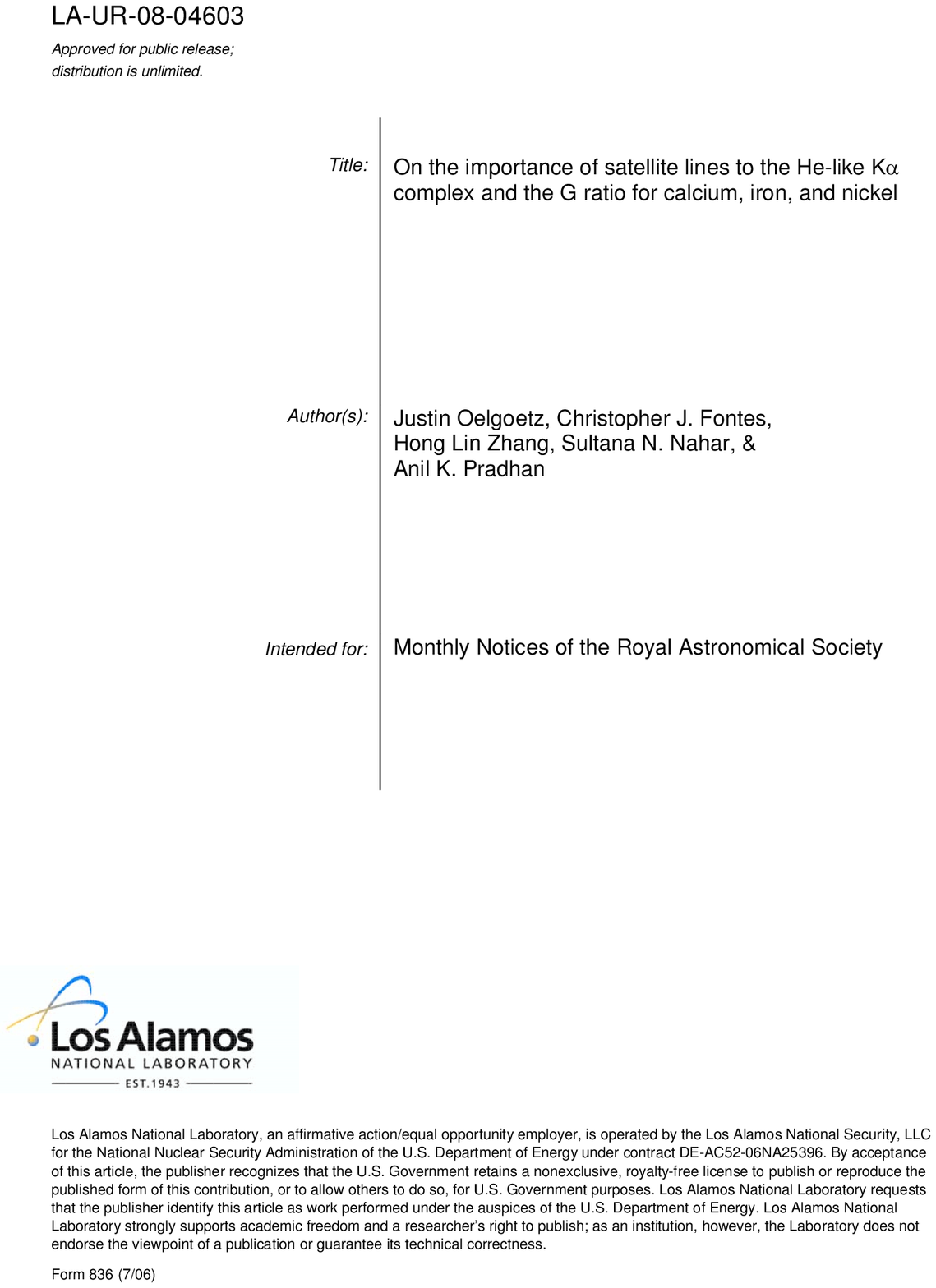}}
\end{center}
\label{firstpage}
\maketitle

\begin{abstract}
New, more detailed calculations of the emission spectra of the He-like K$\alpha$ complex of calcium, iron and nickel have been carried out using data from both distorted-wave and R-matrix calculations.  The value of the $GD$ ratio (an extended definition of the $G$ ratio that accounts for the effect of resolved and unresolved satellite lines) is significantly enhanced at temperatures below the temperature of He-like maximum abundance.  Furthermore it is shown that satellite lines are important contributors to the $GD$ ratio such that $GD/G>1$ at temperatures well above the temperature of maximum abundance.  These new calculations demonstrate, with an improved treatment of the KLn ($\mathrm{n\ge3}$) satellite lines, that K$\alpha$ satellite lines need to be included in models of He-like spectra even at relatively high temperatures.  The excellent agreement between spectra and line ratios calculated from R-matrix and distorted-wave data also confirms the validity of models based on distorted-wave data for highly charged systems, provided the effect of resonances are taken into account as independent processes.
\end{abstract}

\section{introduction}
As described in the literature \citep[e.g.][]{Gabriel-Jordan:1969-helines}, the K$\alpha$ emission of helium-like ions takes place via four lines, with the designations w, x, y and z: $\mathrm{1s2p\;^1P_1\rightarrow1s^2\;^1S_0}$ (w), $\mathrm{1s2p\;^3P_2\rightarrow1s^2\;^1S_0}$ (x), $\mathrm{1s2p\;^3P_1\rightarrow1s^2\;^1S_0}$ (y), $\mathrm{1s2s\;^3S_1\rightarrow1s^2\;^1S_0}$ (z).  From these four lines, line ratios have been investigated for diagnostic purposes.  One of these ratios, $G$, is sensitive to temperature and is defined as
\begin{equation}
G=\frac{I(\mathrm{x})+I(\mathrm{y})+I(\mathrm{z})}{I(\mathrm{w})}\;,\label{GRatio}
\end{equation}
where $I(\mathrm{w})$ is the intensity of the w line in units of number of photons per unit volume per unit time.  For heavier elements, additional lines arising from transitions of the type $\mathrm{1s2lnl' \rightarrow 1s^2nl'}$, where the upper $\mathrm{1s2lnl'}$ states are autoionizing, tend to complicate this simple spectrum \citep[e.g.][]{Edlen-Tyren:1939,Gabriel-Jordan:1969-nature,Gabriel:1972-sats,Mewe-Schrijver:1978stat}.  The KLL satellite lines (which arise from configurations of the type  $\mathrm{1s2l2l'}$) are designated with the letters a--v (see \citealp{Gabriel:1972-sats}; the most recent treatment is given by \citealp{Nahar-Pradhan:2006}).  Higher satellite lines arising from  $\mathrm{1s2lnl'}$, $\mathrm{n}\ge3$, are usually not given separate designations.  Often, astrophysical spectra can not be measured such that these satellite lines are adequately resolved; the result is what appears to be a broadened and redshifted w line according to the intensities of the KLL lines in toto within the K$\alpha$ complex \citep[e.g.][]{Oelgoetz-Pradhan:2001,Hellier-Mukai:2004,Xu-etal:2006}.

\citet{Swartz-Sulkanen:1993} first proposed a method of analysing emission spectra if the resolution was sufficient to resolve the spectra into two ranges, one corresponding to an energy range around the w line, the other including everything redward of this range.  The range about the w line would include not only the w line itself, but also most of the satellite lines arising from the configurations $\mathrm{1s2lnl'}$, $n\ge3$.  The redward range would include the bulk of the KLL satellite lines, in addition to the x, y and z lines.  Thus, they proposed redefining $G$  (referred to as $G_{S\&S}$ below) by taking the integral of the flux redward of some specified boundary line and dividing by the integral of the flux blueward of that same line.

\citet{Bautista-Kallman:2000} considered the same effect by including the intensity of the satellite lines as part of the numerator in their calculation of $G$.  Soon after, \citet{Oelgoetz-Pradhan:2001} proposed a new ratio, $GD$, which included all the KLL satellite lines in the numerator and all the satellite lines arising from higher shells in the denominator.  For some elements, a weak KLL line is present in the area one would associate with the w line; additionally, for most heavier elements, some higher lines (which arise from configurations such as $\mathrm{1s2lnl'}$,~$\mathrm{n\ge3}$) have low enough energies such that they should be included with the x, y and z lines in the numerator \citep{Bely-Dubau-etal:1979a,Bely-Dubau-etal:1979b}.  To improve on these earlier efforts, the $GD$ line ratio is redefined in the current work as
\begin{equation}
GD=\frac{\sum\limits_{E_{K\alpha-\mathrm{min}}<E_s<E_b}I(s)}{\sum\limits_{E_b<E_s<E_{K\alpha-\mathrm{max}}}I(s)}\;, \label{GD}
\end{equation}
where $E_b$ is the boundary line between the two energy ranges, $E_{K\alpha-\mathrm{min}}$ and $E_{K\alpha-\mathrm{max}}$ denote the energy range of the $K\alpha$ complex, and $E_s$ is the energy of a particular line, $s$.  Thus, each sum includes the intensity of each line which has its centroid in the appropriate range.  In the low temperature limit, $G_{S\&S}\approx GD$, but at high temperatures, Doppler broadening will cause the wings of lines near $E_b$ to appear in the other range when computing $G_{S\&S}$.

In addition to being a temperature sensitive diagnostic, the $G$ and $GD$ ratios are also sensitive to the ionization state of the plasma  \citep{Pradhan:1985-cascade,Oelgoetz-Pradhan:2001,Oelgoetz-Pradhan:2004}.  While plasmas out of coronal equilibrium are not considered in this work, the results presented here have direct implications and utility to modeling those systems.

Lastly, it should be noted that some recent work \citep{Rana-etal:2006,Girish-etal:2007} omits the satellite lines from analysis of Fe K$\alpha$ observations on the basis of the argument that \citet{Oelgoetz-Pradhan:2001} showed that the contributions from these lines can be neglected above the temperature of He-like maximum abundance.  While \citet{Oelgoetz-Pradhan:2001} reported that $GD\approx G$ in the range $T_e>3.0\times 10^7$~K \citep[][Fig. 2]{Oelgoetz-Pradhan:2001}, they also showed that the satellites are an important part of the flux in this temperature range \citep[][Fig. 1]{Oelgoetz-Pradhan:2001}.  In these earlier calculations, the contribution of satellite lines to the denominator and numerator of $GD$ effectively cancelled each other out, resulting in $GD\approx G$.  However, the calculations of \citet{Oelgoetz-Pradhan:2001} did not treat the KLn ($\mathrm{n\ge3}$), satellite lines on par with the KLL lines.  Specifically, the KLL lines were treated rigorously according to the method of \citet{Gabriel-Paget:1972}, while the KLn ($\mathrm{n\ge3}$) lines were treated more approximately via a scaling of ratios of autoionization rates.  Additionally, recent work \citep{Oelgoetz-etal:2007-hightg} has shown that the cascade contribution to the recombination rates \citep{Mewe-Schrijver:1978stat} used in \cite{Oelgoetz-Pradhan:2001} diverges from the corresponding contribution calculated with more modern distorted-wave and R-matrix methods at temperatures above the Fe He-like temperature of maximum abundance ($T_e\gtrsim 4\times 10^7$ K).  For these reasons a new study which treats the KLn ($\mathrm{n\ge3}$) satellite lines on par with the KLL satellite lines, and which is also based on more accurate atomic data, is warranted.

\section{Theory}

The present work employed the General Spectral Modeling (GSM) code (\citealp{Oelgoetz-dissertation}, see also \citealp{Oelgoetz-etal:2007-hightg,Oelgoetz-etal:2007-highNe-R}).  GSM is based on the ground-state-only quasi-static approximation \citep[e.g.][]{Bates-Kingston-McWhirter:1962-thin}, a common method for modelling low density plasmas such as found in astrophysics, which assumes that the ionisation balance portion of the model can be separated from a determination of excited-state populations. The rationale for this approximation is two fold: first, the times scales for ionisation and recombination are much longer than the time scales for processes inside an ionisation stage, and second, the populations of the excited states have a negligible effect on ionisation and recombination. \citep[See][for a discussion of the validity of this approximation.]{Oelgoetz-etal:2007-highNe-R}  Thus, the first step in a GSM calculation is to solve the coupled set of ionisation balance equations given by
\begin{eqnarray}
\frac{dX_l}{dt}&=&N_e(X_{l+1}\alpha_{l+1 \rightarrow l}(T_e)+X_{l-1}C_{l-1 \rightarrow l}(T_e)) \nonumber \\ && + N_e^2(X_{l+1}\beta_{l+1 \rightarrow l}(T_e) - X_{l}\beta_{l \rightarrow l-1}(T_e)) \nonumber \\ && - X_{l}N_e(\alpha_{l \rightarrow l-1}(T_e) + C_{l \rightarrow l+1}(T_e)) \label{ionbaleqn} \;,
\end{eqnarray}
where $X_i$ is the total population in the $i^{th}$ ionisation stage, $N_e$ the electron number density, $T_e$ is the electron temperature, $C$ is a bulk collisional ionisation rate coefficient, $\beta$ a bulk 3-body recombination rate coefficient, and $\alpha$ a bulk recombination rate coefficient (which includes radiative and dielectronic recombination).  In general, photoionization and stimulated recombination are included as well, but as this work considers only collisional plasmas, the rate coefficients associated with these processes have been omitted from equation (\ref{ionbaleqn}).

Once the values of $X_l$ have been determined, the ground-state-only quasi-static approximation then allows one to solve for the excited-state populations in a given ionisation stage, with the approximation of treating the ionisation stages adjacent to the ionisation stage of interest as being entirely in the ground state.  As the total population in the ionisation stage of interest and the two adjacent ionisation stages are known, the excited-state populations can be determined by solving a modified version of the full set of collisional-radiative equations given by
\begin{eqnarray}
\frac{dN_{l,j}}{dt} & = &N_e\left( \sum_{i(i\ne j)}(N_{l,i}q^{\mathrm{eff}}_{i \rightarrow j}(T_e)-N_{l,j}q^{\mathrm{eff}}_{j \rightarrow i}(T_e)) \right.\nonumber \\
&+& X_{l+1}C^{\mathrm{eff}}_{l+1,1 \rightarrow l,j}(T_e)-N_{l,j}\sum_{i}C^{\mathrm{eff}}_{l,j \rightarrow l-1,i}(T_e) \nonumber \\ 
&+&\left.  X_{l-1}\alpha^{\mathrm{eff}}_{l-1,1 \rightarrow l,j}(T_e) -N_{l,j}\sum_{i}\alpha^{\mathrm{eff}}_{l,j \rightarrow l+1,i}(T_e) \right)\nonumber \\ 
&+&\sum_{i(i>j)}N_{l,i}A^{\mathrm{eff}}_{i \rightarrow j}+N_e^2\bigg(X_{l-1}\beta^{\mathrm{eff}}_{l-1,1 \rightarrow l,j}(T_e) \nonumber \\
&-&N_{l,j}\sum_{i}\beta^{\mathrm{eff}}_{l,j \rightarrow l+1,i}(T_e)\bigg)-N_{l,j}\sum_{i(i<j)}A^{\mathrm{eff}}_{j \rightarrow i} \nonumber \\
&+&N_{p}\left(\sum_{i(i\ne j)}(N_{l,i}q^{\mathrm{p-eff}}_{i \rightarrow j}(T_e)-N_{l,j}q^{\mathrm{p-eff}}_{j \rightarrow i}(T_e))\right) \nonumber \\
&+&N_{\alpha}\left(\sum_{i(i\ne j)}(N_{l,i}q^{\mathrm{\alpha-eff}}_{i \rightarrow j}(T_e)-N_{l,j}q^{\mathrm{\alpha-eff}}_{j \rightarrow i}(T_e))\right) \nonumber \\
&-&N_{l,j}\sum_{i} R^{\mathrm{AI-eff}}_{l,j \rightarrow l-1,i}
 \;,\label{excitedformula} \\
X_l &=& \sum_j N_{l,j}\; \label{matcons},
\end{eqnarray}
where the variables are defined more or less as before; $N_{l,j}$ is the population in the $j^{th}$ state of the $l^{th}$ ionisation stage, $q^{\mathrm{eff}}$ is an electron-impact (de-)excitation effective rate coefficient, $q^{\mathrm{p-eff}}$ is a proton-impact (de-)excitation effective rate coefficient, $q^{\mathrm{\alpha-eff}}$ is an alpha-particle-impact (de-)excitation effective rate coefficient, $R^{\mathrm{AI-eff}}$ is an effective autoionization rate coefficient, and $A^{\mathrm{eff}}$ is an effective radiative decay rate.  Here we have used the ``eff'' superscript to denote the possible use of effective rate coefficients since GSM offers the option of treating some of the excited states as statistical conduits (using branching ratios) and others explicitly \citep[e.g.][]{Oelgoetz-dissertation,Oelgoetz-etal:2007-hightg,Oelgoetz-etal:2007-highNe-R}.  Explicit states are those that appear in the set of coupled equations presented in equations (\ref{excitedformula}) and (\ref{matcons}).  When all states within an ionisation stage are treated explicitly, the ``eff'' superscript is not necessary since all of the rate coefficients represent direct processes only.  When the statistical treatment is employed, the rate coefficients associated with the processes passing through statistical states are combined with the direct rate coefficients between explicit levels by summing over all the indirect paths through the statistical states.  This process is simplified by the use of the {\it collisionless transition matrix} (CTM), $\mathbf{T}_{m \rightarrow j}$, which can be thought of as the probability that an ion in statistical state $m$ will end up in an explicit state $j$, assuming that the time scale for collisions is very long when compared to the time scale for the spontaneous processes of autoionization and radiative decay.  If $Q$ represents the set of explicit states, and $i$ a state such that $i\not\in Q$, the CTM can be defined using the recursive expression
\begin{eqnarray}
\mathbf{T}_{i\rightarrow j}&=&\sum\limits_{k\not\in Q \atop (E_i\ge E_k\ge E_j)}\frac{\Gamma_{i\rightarrow k}}{\sum\limits_lA_{i\rightarrow l}+\sum\limits_m R^\mathrm{AI}_{i \rightarrow m}}\mathbf{T}_{k\rightarrow j}\nonumber \\
&+&\frac{\Gamma_{i\rightarrow j}}{\sum\limits_lA_{i\rightarrow l}+\sum\limits_m R^\mathrm{AI}_{i\rightarrow m}}\;,
\end{eqnarray}
where $\Gamma_{i\rightarrow k}$ is the appropriate type of spontaneous rate (either radiative decay or autoionization) to connect states $i$ and $k$ .  It should be noted that if $i\in Q$ the CTM is not meaningful, and as such is defined to be zero.  The effective rate coefficient is then calculated by summing the direct rates, and the fraction (as determined by the CTM) of each indirect rate which contributes to an effective rate.  For example, effective recombination (RR+DR) rate coefficients are calculated as 
\begin{eqnarray}
\alpha^{\mathrm{eff}}_{j \rightarrow i}(T_e) &=& \alpha^\mathrm{RR}_{j \rightarrow i}(T_e) + D^\mathrm{DC}_{j \rightarrow i}(T_e) \nonumber \\ &+& \sum_{m\atop(E_m>E_i)}\alpha^\mathrm{RR}_{j \rightarrow m}(T_e)\mathbf{T}_{m \rightarrow i} \nonumber \\ &+& \sum_{m\atop{(E_m>0)\atop{(E_m>E_i)\atop(E_m>E_j)}}}D^\mathrm{DC}_{j \rightarrow m}(T_e)\mathbf{T}_{m \rightarrow i} \label{rceq}\;,
\end{eqnarray}
where $D^\mathrm{DC}$ is a dielectronic capture rate and the sums take into account both radiative recombination and dielectronic capture followed by radiative cascade.  It should be noted that there are terms in these sums that would be represented by explicit resonances in R-matrix cross sections.  Such an approach allows for the inclusion of resonances when perturbative (e.g. distorted-wave) cross sections are employed.  This approach is sometimes referred to as the independent-process, isolated-resonance (IPIR) method \citep[see][]{Bates-Dalgarno:1962,Gabriel-Paget:1972,Cowan-twostep:1980,Badnell-etal:1993-IPIRDW-breakdown}.  In calculations that consider R-matrix data, care must be taken to exclude these terms from the summations in equation (\ref{rceq}) in order to avoid double counting the resonance contributions.

Once excited-state populations have been calculated, the intensity of each line in the spectral region of interest is calculated according to
\begin{equation}
I(l,j\rightarrow l,k)=N_{l,j}A_{j \rightarrow k}\;.
\end{equation}
Each line is then given a line shape corresponding to a thermal Doppler-broadened Gaussian profile.  The total spectrum (or emissivity), $S$, for a given photon energy, $h\nu$, can be expressed as
\begin{equation}
S(h\nu)=h\nu \sum_s I_s\frac{c}{\Delta E_s}\sqrt{\frac{m_i}{2\pi kT_i}}\,e^{\frac{m_ic^2(h\nu-\Delta E_s)^2}{2(\Delta E_s)^2kT_i}}\;,
\end{equation}
where $S$ is in units of energy per unit volume per unit time per energy interval, $s$ ranges over the set of all included transitions in the desired energy range, $\Delta E_s$ is the transition energy associated with a given line, and the ion temperature, $T_i$, is taken to be equal to the electron temperature.

\section{Computations}
As this work is concerned with steady-state plasmas, the solution to the coupled set of ionisation balance equations, Eq. (\ref{ionbaleqn}), were taken to be those of \citet{Mazzotta-etal:1998} for all three elements (Ca, Fe, and Ni) considered in this work.  Furthermore, as the cases considered are well within the low density limit ($N_e=10^{10}$ cm$^{-3}$), the approximation \citet{Mazzotta-etal:1998} made in neglecting three-body recombination is valid.

The present work considered multiple classes of models for each of the three elements.  Each model contains a different set of detailed atomic data.  The first class, composed mostly of distorted-wave (DW) data (and denoted by Ni:DW, Fe:DW, and Ca:DW for the three elements), uses a set of data calculated entirely by the Los Alamos suite of atomic physics codes \citep[e.g.][]{Abdallah-etal:1994,Abdallah-etal:2001}.  The CATS code was used to calculate the wave functions, energies, and dipole allowed radiative decay rates for all fine-structure levels arising from the configurations $\mathrm{nl}$, $\mathrm{1snl}$, $\mathrm{2lnl'}$, $\mathrm{1s^2nl}$, $\mathrm{1s2lnl'}$, and $\mathrm{1s3lnl'}$ with $\mathrm{n}\le 10$ and $\mathrm{l}\le g$, which span the H-like, He-like, and Li-like ionisation stages.  The GIPPER code was used to calculate all autoionization rates and photoionization cross sections in the distorted-wave approximation, as well as collisional ionisation cross sections using a scaled-hydrogenic approximation which has been shown to agree well with distorted-wave results for highly charged systems.  Distorted-wave cross sections for all electron-impact excitation transitions out of the lowest seven levels of the helium-like ionisation stage, as well as the $\mathrm{1s^22l}$ complex of the Li-like ionisation stage were calculated with the ACE code.  Cross sections for the remaining electron-impact excitation transitions were computed in the more approximate plane-wave Born approximation.  Lastly, the non-dipole $A$ values that give rise to the x and z lines, as well as a two-photon decay rate from $\mathrm{1s2s\;^1S_0 \rightarrow 1s^2\;^1S_0}$ used in obtaining the populations from equations (\ref{excitedformula}) and (\ref{matcons}), were obtained from \citet{Mewe-Schrijver:1978stat}.  Proton- and alpha-particle-impact excitation rates between the He-like $\mathrm{1s2l}$ levels were also taken from \citet{Mewe-Schrijver:1978stat}.  The protons and alpha particles were taken to have the same temperature as the electrons, and to have densities of 0.77 and 0.115 times the electron density respectively ($N_p=0.77 N_e$, $N_\alpha=0.115N_e$).  The CATS level energies for the lowest seven levels of the He-like ionisation stage and the lowest three levels of the Li-like ionisation stage were replaced by values taken from the NIST Atomic Spectra database \citep{NISTDB}, as were the energies for the KLL autoionizing levels for Li-like Ni and Fe.  As the NIST database does not contain complete information for the autoionizing KLL levels of Ca, the level energies calculated by CATS were retained for all Ca autoionizing states.  All of the fine-structure levels arising from the $\mathrm{1s}$, $\mathrm{1s^2}$, $\mathrm{1s2l}$, $\mathrm{1s^22l}$, and $\mathrm{1s2lnl'}$ configurations with $\mathrm{n}\le 10$ and $\mathrm{l}\le g$ were treated explicitly when solving for the excited-state populations appearing in equations (\ref{excitedformula}) and (\ref{matcons}).

In the second class of models the electron-impact excitation, radiative decay, and both radiative and dielectronic recombination data in the DW model are replaced with data calculated using R-matrix (RM) methods, where such data are publicly available.  For Ni, the radiative decay rates of \citet{Nahar-NiXXVII-Avalues} and the unified recombination rates of \citet{Nahar:2005} were used to create the Ni:RM data set.  As for Fe, two sets of R-matrix electron-impact excitation rates are available and are considered here.  The first set, Fe:RM, includes the electron-impact excitation collision strengths of \citet{Pradhan:1985-eie}, a subset of the radiative decay rates of \citet{Nahar-Pradhan:1999} (where the initial state is a fine-structure level arising from the configurations $\mathrm{1snl}$ where $\mathrm{n}\le 4$, $\mathrm{l}\le f$ or $5 \le \mathrm{n}\le 10$, $\mathrm{l}\le p$) and the corresponding subset of the unified recombination rates of \citet{Nahar-etal:2001-FeXXIV-FeXXVrc} ($\mathrm{1s}$ $\mathrm{^2S_{1/2}}$ recombining into all fine-structure levels arising from $\mathrm{1snl}$ where $\mathrm{n}\le 4$, $\mathrm{l}\le f$ or $5 \le \mathrm{n}\le 10$, $\mathrm{l}\le p$).  The second set, Fe:RM2, is identical to Fe:RM except that it uses the electron-impact excitation collision strengths of \citet{Whiteford-etal:2001}.  Lastly, one R-matrix type model is considered for Ca, Ca:RM, which also incorporates the electron-impact excitation data of \citet{Pradhan:1985-eie}.

The last class of models is an expansion of the second class by also incorporating autoionization rates calculated from recombination cross sections \citep[e.g.][]{Nahar-sat-rc2}.  Specifically, \citet{Nahar-sat-rc2} provided this type of data for Fe and Ni.  These data have been combined with the Fe:RM and Ni:RM sets to make the Fe:RM+ and Ni:RM+ sets.  The Fe:RM2 data set, which incorporates the collision strengths of \citet{Whiteford-etal:2001}, has not been expanded into a Fe:RM2+ data set due to the good agreement (which is shown in the following section) between the Fe:RM and Fe:RM2 data set.  As no data of this type are yet available for Ca, no model of this class is considered for Ca.

In addition to constructing the models, the boundary line between the high energy and low energy section of each spectrum had to be chosen.  As pointed out by \citet{Swartz-Sulkanen:1993} there is an energy gap that forms between the w line (and the satellite lines that blend with it) and the rest of the spectrum.  This gap was found by inspection, and the boundary energy, $E_b$, was chosen to be 3895 eV, 6690 eV, and 7794 eV for Ca, Fe, and Ni respectively.  

\section{Results}

Figs. \ref{Ca-GD}--\ref{Ni-GD} display the calculated values of the $G$ and $GD$ ratios as a function of temperature for each of the models, along with plots of certain ratios that help to illustrate where the differences occur.

\begin{figure*}
\vspace{\figspace}
\begin{center}
\resizebox{!}{\figheight}{\includegraphics{fig1.eps}}
\end{center}
\caption{{\bf  Top panel}: $GD$ and $G$ ratios as a function of electron temperature for the Ca:DW and Ca:RM data sets.  The temperature of maximum abundance (T$_m\sim1.5\times10^7$ K) has been indicated with an arrow.  {\bf Middle panel}: Ratios of $G$ and $GD$ ratios computed from the various data sets. {\bf Bottom panel}: Ratios of $GD$ to $G$ for each of the data sets.  {\bf Insert on bottom panel}: The same ratios of $GD$ to $G$ as the bottom panel, but plotted with a linear scale on the y axis to better resolve the behaviour.\label{Ca-GD}}
\end{figure*}

\begin{figure*}
\vspace{\figspace}
\begin{center}
\resizebox{!}{\figheight}{\includegraphics{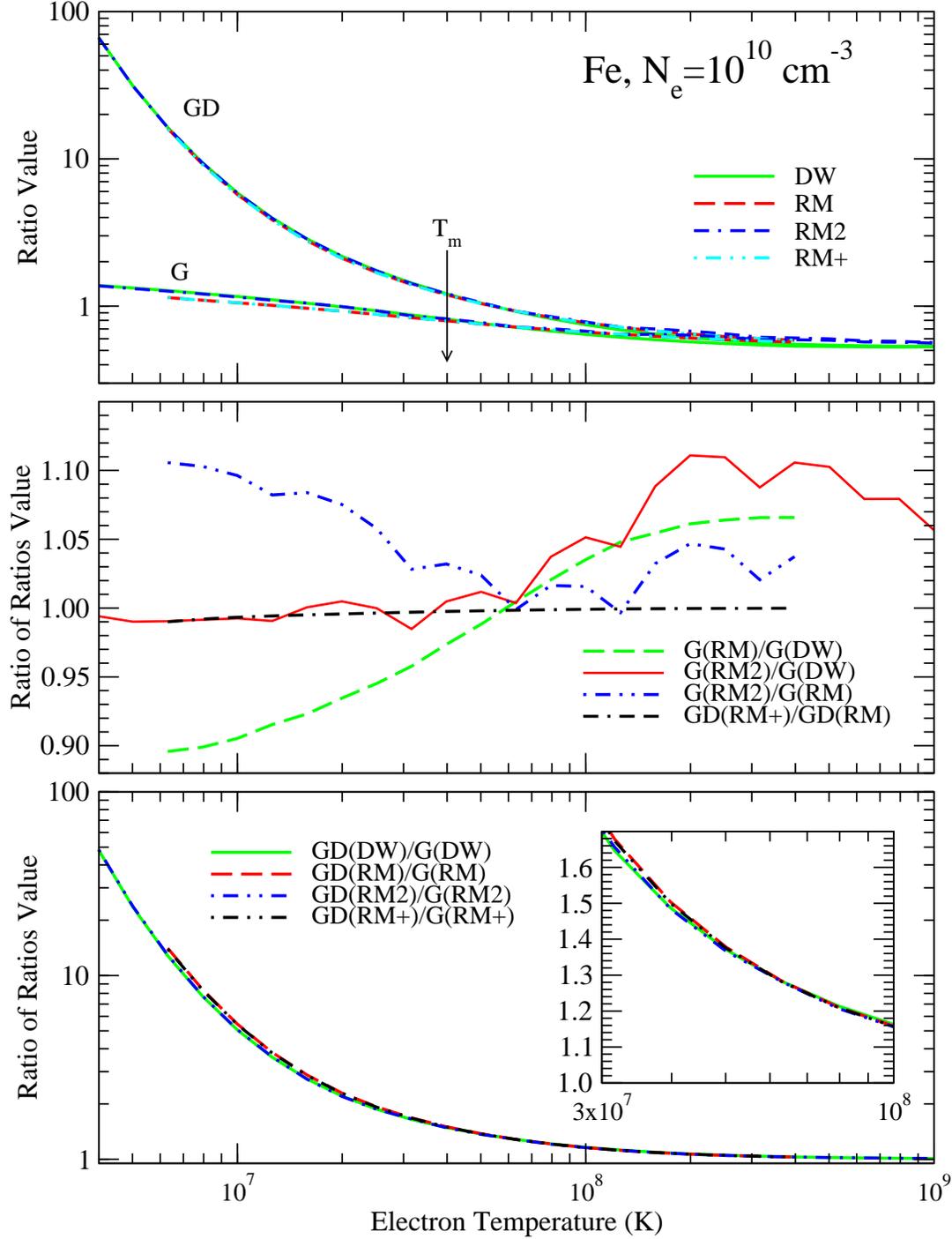}}
\end{center}
\caption{{\bf  Top panel}: $GD$ and $G$ ratios as a function of electron temperature for the Fe:DW, Fe:RM, Fe:RM2, and Fe:RM+ data sets.  The temperature of maximum abundance (T$_m\sim4\times10^7$ K) has been indicated with an arrow. {\bf Middle panel}: Ratios of $G$ and $GD$ ratios computed from the various data sets. The humps in curves that involve the RM2 data set are due to interpolation on the rates of \citet{Whiteford-etal:2001}.  {\bf Bottom panel}: Ratios of $GD$ to $G$ for each of the data sets.  {\bf Insert on bottom panel}: The same ratios of $GD$ to $G$ as the bottom panel, but plotted with a linear scale on the y axis to better resolve the behaviour.\label{Fe-GD}}
\end{figure*}

\begin{figure*}
\vspace{\figspace}
\begin{center}
\resizebox{!}{\figheight}{\includegraphics{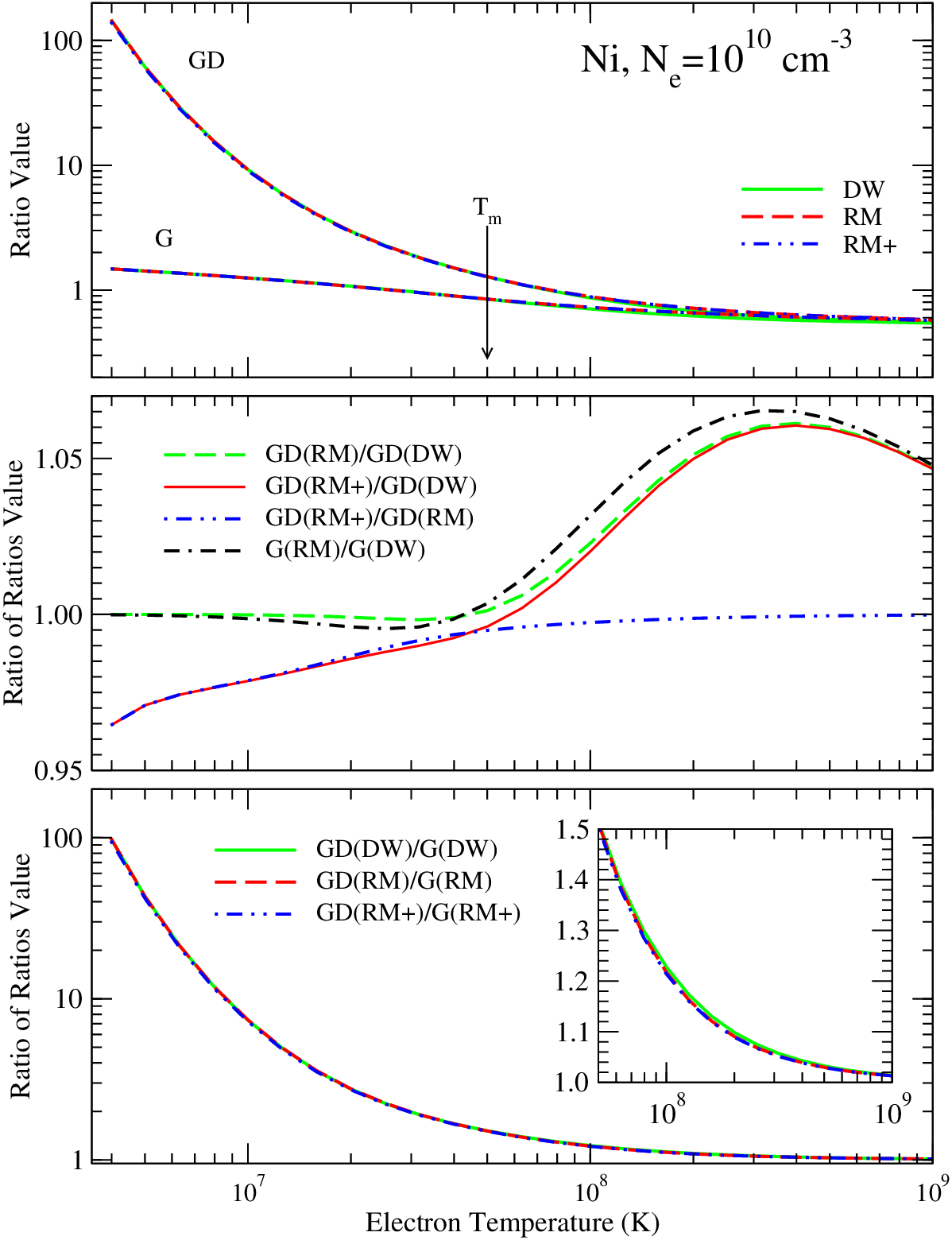}}
\end{center}
\caption{{\bf  Top panel}: $GD$ and $G$ ratios as a function of electron temperature for the Ni:DW, Ni:RM, and Ni:RM+ data sets.  The temperature of maximum abundance (T$_m\sim5\times10^7$ K) has been indicated with an arrow. {\bf Middle panel}: Ratios of $G$ and $GD$ ratios computed from the various data sets. {\bf Bottom panel}: Ratios of $GD$ to $G$ for each of the data sets.  {\bf Insert on bottom panel}: The same ratios of $GD$ to $G$ as the bottom panel, but plotted with a linear scale on the y axis to better resolve the behaviour.\label{Ni-GD}}
\end{figure*}

Overall, the present calculations predict $GD$ ratios that are significantly higher than the corresponding $G$ ratios for all the models that are considered.  This behaviour is in qualitative agreement with previous studies \citep{Swartz-Sulkanen:1993,Bautista-Kallman:2000,Oelgoetz-Pradhan:2001}; it should be noted that this more detailed study predicts a significantly greater value of $GD$ below the temperature of maximum abundance than any of the previous studies.  Additionally, the impact of satellite lines on the $GD$ ratio keeps the $GD$/$G$ ratio greater than one over a much broader range than shown in the study of \citet{Oelgoetz-Pradhan:2001}.  The principal reason for this behaviour is that the more approximate treatment of KLM and higher lines in \citet{Oelgoetz-Pradhan:2001} appears to overestimate their importance, especially at higher temperatures \citep[see][Fig. 3]{Oelgoetz-Pradhan:2001}.  This overestimation leads to a cancelling effect, whereby the KLM and higher lines in the denominator of $GD$ cancel out the effect of the KLL satellite lines in the numerator.  Additionally, the present calculations allow the KLM and higher satellite lines to be included within the energy range they actually fall, which is in the redward section (i.e. the numerator of $GD$---with the x, y, and z lines) for some of the higher satellite lines.  Thus, the satellite lines in these new calculations have an impact on the line ratio $GD$ at temperatures well above the temperature of maximum abundance for the He-like ionisation stage.  One practical consequence of this last statement is that essentially any spectral analysis of the He-like K$\alpha$ lines requires the satellite lines to be treated in a detailed manner (unless the measured spectra are sufficiently well resolved so that the satellite lines can be readily distinguished).  Due to the level of detail and improved atomic data included in the present calculations, they are expected to be a significant improvement over previous work.  

While there are differences between the $G$ ratios, as well as the $GD$ ratios, predicted by each of the data sets, these differences are all less than 15 percent, which is within the typical 10--20 percent uncertainty reported for the R-matrix data \citep[e.g.][]{Nahar-etal:2001-FeXXIV-FeXXVrc,Pradhan:1985-eie,Whiteford-etal:2001}.  In order to understand these differences, spectra were examined for a wide range of temperatures.  In general, spectra for all the elements and models considered were found to be in excellent agreement with each other, even when comparing results obtained from RM and DW data sets.  The differences were all less than 12 percent for strong lines, which include the w, x, y, and z lines, as well as most of the satellite lines.  There were larger differences (up to $\sim$50 percent) for some weak but barely visible satellite lines (like c), and even larger differences (up to $\sim$150 percent) for some weaker satellite lines that do not contribute in any appreciable manner to the spectra.  These larger differences have very little impact on the spectra or the line ratios as the corresponding lines are quite weak. 
 
\begin{figure*}
\vspace{\figspace}
\begin{center}
\resizebox{!}{\sfigheight}{\includegraphics{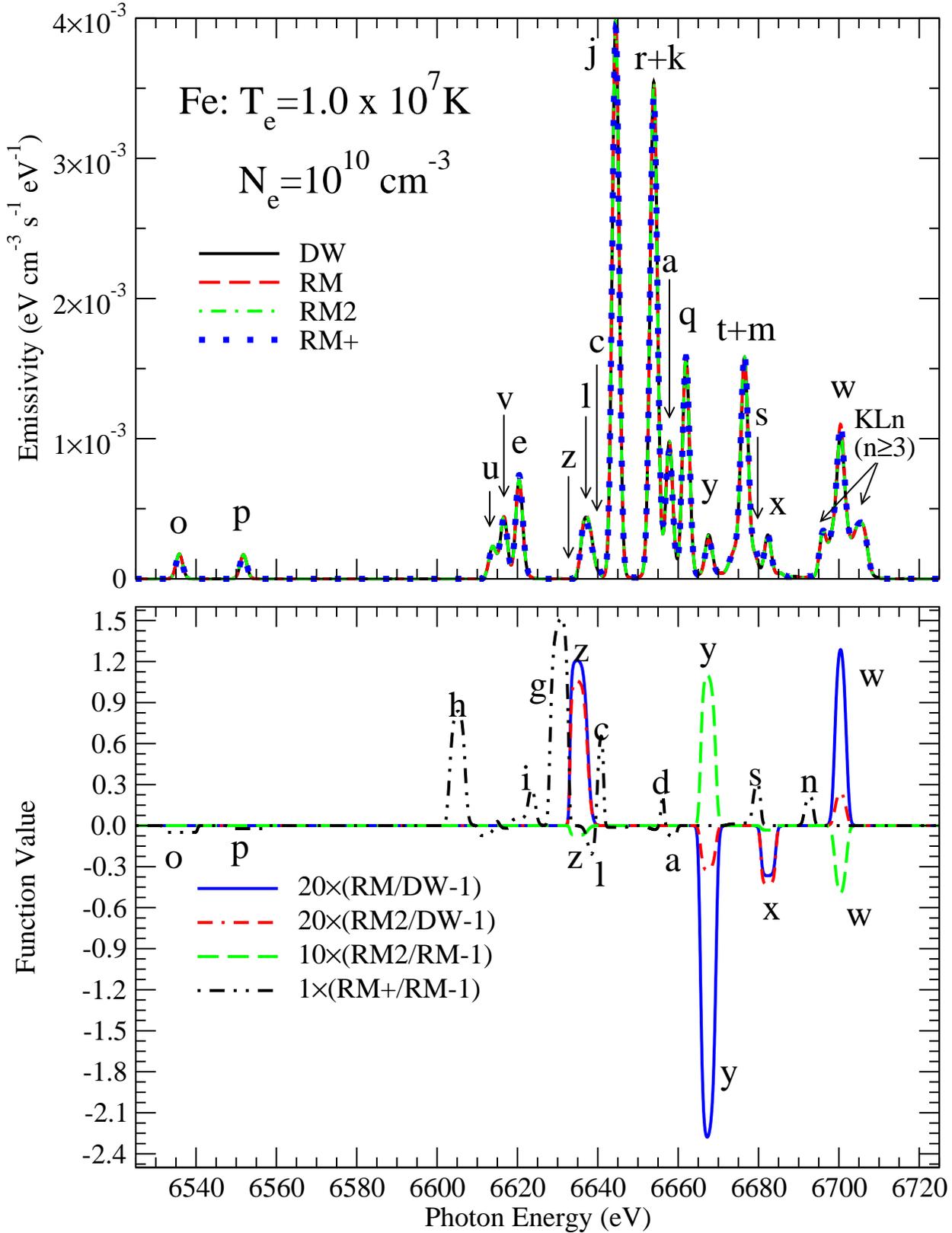}}
\end{center}
\caption{{\bf Top panel}: Calculated Fe spectra at T$_e=1.0\times 10^7$ K, N$_e=10^{10}$ cm$^{-3}$ for each of the Fe data sets. {\bf Bottom panel}: Functions of the spectra presented for comparison.  Of particular note is the excellent agreement between all spectra, and that the largest uncertainties are in very weak satellite lines.  While the satellite lines in the redward portion of the spectrum are dominated by the KLL lines (a--v), the KLn (n$\ge$3) lines are significant enough to blend with some of the KLL satellite lines and alter the shape of the feature, even if they do not appear as separate features like they do near the w line.\label{Fe-1}}
\end{figure*}

\begin{figure*}
\vspace{\figspace}
\begin{center}
\resizebox{!}{\sfigheight}{\includegraphics{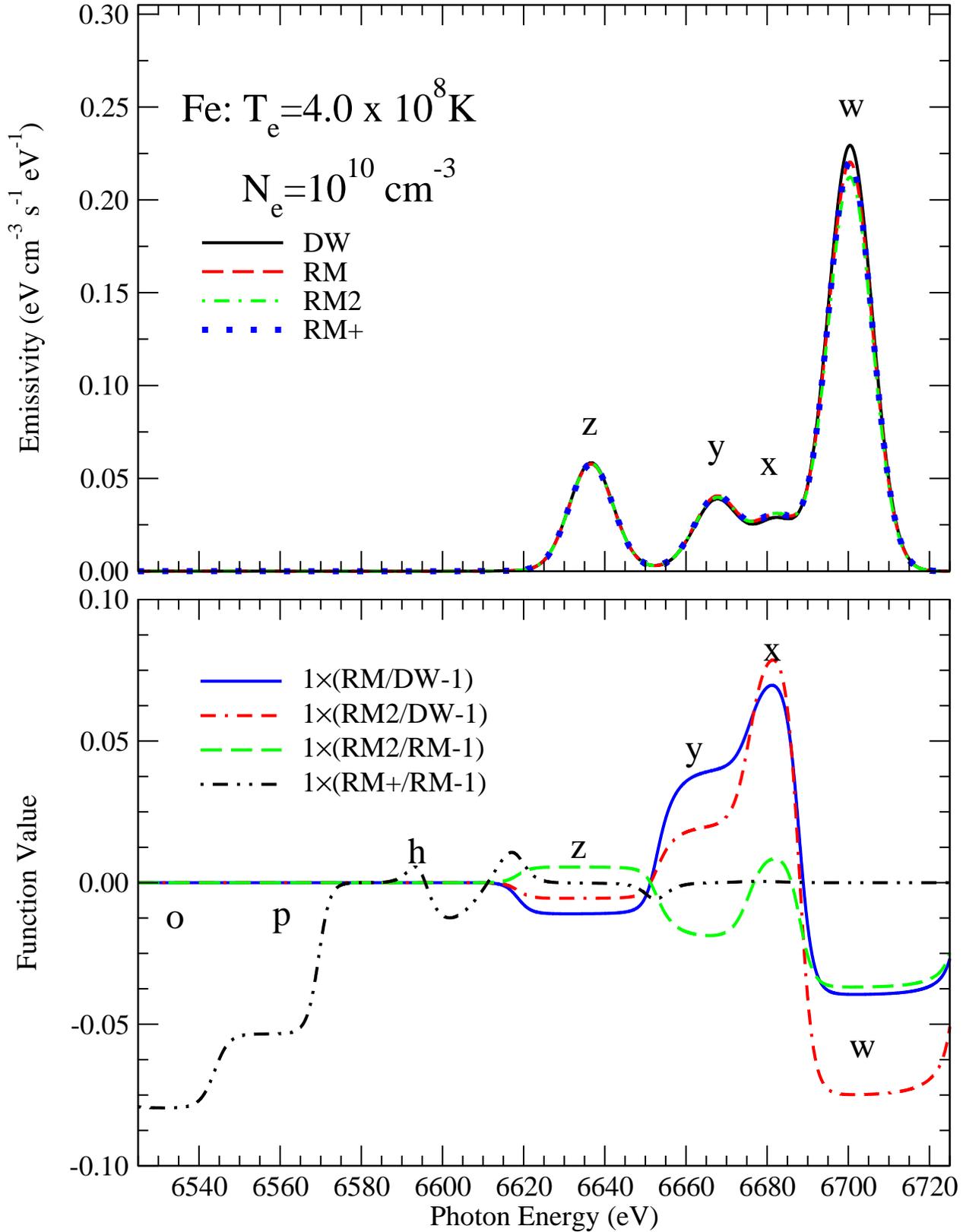}}
\end{center}
\caption{{\bf Top panel}: Calculated Fe spectra at T$_e=4.0\times 10^8$ K, N$_e=10^{10}$ cm$^{-3}$ for each of the Fe data sets. {\bf Bottom panel}: Functions of the spectra presented for comparison.  Again there is excellent agreement between all spectra.  The largest uncertainties are in very weak satellite lines, as well as the w and x lines.\label{Fe-2}}
\end{figure*}

Two sample spectra for Fe, for which the disagreement in the ratios was among the largest, are presented in Figs.~\ref{Fe-1} and \ref{Fe-2}.  As illustrated in the upper panel of Fig.~\ref{Fe-1}, at an electron temperature of $10^7$ K, the overall agreement between the spectra computed with the various models is excellent. The data in the bottom panel of Fig.~\ref{Fe-1} indicate more precisely where the largest discrepancies occur.  One observes that the use of R-matrix data results in an increase of the z line and a decrease in the x line relative to the distorted-wave model.  Additionally, the Fe:RM data set predicts a decrease in the y line, and an increase in the w line relative to the distorted-wave model; the Fe:RM2 data set predicts the same changes, but to a lesser extent.  From this inspection one can conclude that the agreement between the $G$ ratios calculated from the Fe:DW and Fe:RM2 data sets is fortuitous because of a cancellation in the quantities that comprise the numerator and denominator of that ratio.  On the other hand, the decrease in the x and y lines predicted by the Fe:RM versus the Fe:DW data set are larger than the corresponding increase in the z line.  This overall reduction in the numerator of the $G$ ratio, when coupled with the increase in the w line between the Fe:RM and Fe:DW data sets, results in the reduced $G$ ratio calculated from the Fe:RM model at low temperatures.

Fig. \ref{Fe-2}, which displays spectra at a much higher electron temperature of $4\times10^8$ K (which is approximately ten times higher than the temperature of maximum abundance for He-like Fe) again shows excellent agreement.  An analysis of the bottom panel of Fig. \ref{Fe-2} shows that both R-matrix data sets predict higher x and y lines, and a decreased w line, relative to the distorted-wave results.  The net result of these differences is the increased $G$ and $GD$ ratios displayed in Fig. \ref{Fe-GD}.  Separate calculations (not shown) indicate that the increase in the x line is due to slightly higher R-matrix recombination rates rather than to sensitivity to the electron-impact excitation rates.  This populating mechanism for the x line is consistent with the typical viewpoint in the literature \citep[e.g.][]{Pradhan:1985-cascade}.  The y line, on the other hand, is sensitive to both electron-impact excitation and recombination rates at this high temperature; for this case the recombination rates are dominant in determining the population of the excited state, but the excitation rate is non-negligible as y is an intercombination line.
While this temperature ($T_e=4.0\times 10^8$ K) is above the peak of the DR hump \citep[see][Fig. 5]{Nahar-etal:2001-FeXXIV-FeXXVrc}, it is still in a range where the resonances of the R-matrix cross section are important to the recombination rate.  The high-temperature differences observed for the Ni $G$ and $GD$ ratios (Fig. \ref{Ni-GD}), for which only the recombination rates were changed among the various models, have a similar explanation.  Additionally, separate calculations (not shown) indicate that the differences in the w line are primarily due to differences in the electron-impact excitation data.  The importance of excitation over recombination as a populating mechanism  of the w line is expected since this transition is dipole allowed \citep[e.g.][]{Pradhan:1985-cascade}.  The net effect of these differences is the increase in the R-matrix $G$ and $GD$ ratios which is observed above the temperature of maximum abundance in Fig. \ref{Fe-GD}.  

Despite the subtle differences in the spectra presented above, we emphasise that the discrepancies in the important lines are well within the uncertainties (20 percent) usually cited for R-matrix data.  The disagreement in these spectra were among the largest seen in this study, which speaks to the excellent overall agreement between the RM and DW models.

Lastly, it should be noted that the line positions for the KLM and higher satellite lines are a significant source of uncertainty in these calculations.  While the accuracy of the line positions is estimated to be $\sim$0.1 percent, a shift of that size could impact the spectra significantly by causing some of the strong KLM lines, which blend with the w line in this present work, to move sufficiently far such that they should be considered with the bulk of the KLL lines in the numerator of $GD$.  This fact is underscored by the appearance of KLM and higher lines $\sim$7 eV blueward of the w line in Fig. \ref{Fe-1}, when they should instead converge upon the w line.  If some of these higher lying satellite lines do in fact blend with the x line, the impact would be a corresponding increase in the $GD$ ratio.

\section{Conclusions}

New, more detailed calculations of the emission spectra of the He-like K$\alpha$ complex of calcium, iron, and nickel have been carried out using atomic data from both distorted-wave and R-matrix calculations.  Spectra from these calculations are in excellent agreement, and demonstrate that satellite lines are important to both the spectra and the $GD$ ratio across a wide temperature range that includes temperatures significantly above the temperature of maximum abundance for the He-like ionisation stage.  A major conclusion of this work is the need to include satellite lines in the diagnosis of He-like K$\alpha$ spectra of iron peak elements in low density, collisional (coronal) plasmas, even at temperatures well above the temperature of maximum abundance.  When the satellite lines are appropriately taken into account, the $GD$ ratio remains an excellent potential temperature diagnostic.

Another important application of the results presented herein, is in the well-known application of the $G$ or $GD$ ratio to ascertain the ionization state of a plasma. As shown in Figs. \ref{Ca-GD}--\ref{Ni-GD}, the $GD$ ratio is far more sensitive to the ionization state at $T < T_m$ than the $G$ ratio, by as much as a factor of 100. Therefore, it is imperative to calculate the $GD$ values as precisely as possible at temperatures where the dielectronic satellite intensities are rapidly varying. Such conditions are known to occur in plasmas which are not in coronal equilibrium, as discussed by \citet{Pradhan:1985-cascade,Oelgoetz-Pradhan:2001,Oelgoetz-Pradhan:2004}.  Furthermore, it should be noted that, while this work does not consider the effect of satellite lines on the density sensitive diagnostic ratio $R$ ($R = (I (x) + I (y ))/I (z)$), the effect of these lines is significant enough that they would need to be taken into account under conditions where $R$ is used.  This inclusion is warranted due to the manifestation of the satellites embedded within the K$\alpha$ complex, and in many cases blended with the principal lines x, y and z.

The excellent agreement between the spectra produced from R-matrix and distorted-wave data used in the models presented in this work bolsters confidence in both data sets.  Any disagreement between the two sets of spectra would have indicated an error in the fundamental atomic data because the IPIR approach has been shown to give good agreement with close-coupling approaches when producing the fundamental rate coefficients \citep[e.g.][]{Bates-Dalgarno:1962,Gabriel-Paget:1972,Cowan-twostep:1980,Badnell-etal:1993-IPIRDW-breakdown}. The present work provides a more stringent test of this assumption by including those rate coefficients in a fully integrated spectral calculation that takes into account several ion stages and includes the coupling between all of the important atomic processes.

The good agreement observed in this work reaffirms the fact that in highly charged systems, models based on data calculated from computationally less expensive distorted-wave methods can reproduce the results of models based on R-matrix data if the effect of resonances are taken into account as independent processes. This behaviour, however, is not expected to remain true for all conditions, especially when near neutral systems are prevalent.

The results presented in this paper should be applicable to high-energy and high-resolution X-ray spectroscopy of laboratory and astrophysical plasmas.  Astrophysical observations of the K$\alpha$ complex of high-Z ions, particularly the 6.6--6.7 keV range of the Fe K$\alpha$, were expected to be made by the high-resolution X-ray satellite Suzaku, but could not be performed due to instrument failure. It is, however, expected that these calculated results would be valuable in future X-rays missions such as the recently planned joint ESA-NASA International X-ray Observatory.

\section{acknowledgments}
This work was partially conducted under the auspices of the United States Department of Energy at Los Alamos National Laboratory.  Much of the development of GSM was also done at the Ohio Supercomputer Center in Columbus, Ohio (USA).  The work by the OSU group (SNN, AKP) was partially supported by a grant from the NASA Astrophysical Theory Program.

\label{lastpage}
\end{document}